\documentclass[]{aa}
\usepackage{epsfig}
\usepackage{psfig}
\usepackage{graphics}
\usepackage{latexsym}
\begin{document}
\title{Location of narrowband spikes in solar flares}
\author{A.O. Benz\inst{1} \and P. Saint-Hilaire\inst{1,2} \and N. Vilmer\inst{3}}
\offprints{A. O. Benz}
\mail{benz@astro.phys.ethz.ch}
\institute{Institute of Astronomy, ETH Zentrum, CH-8092 Zurich, Switzerland \and Paul Scherrer Institute, CH-5232 Villigen PSI, Switzerland \and
Observatoire de Paris, Section de Meudon, DASOP \& CNRS UMR 8645, 92195 Meudon, France}
\date{}
\titlerunning{Narrowband spikes in solar flares}
\authorrunning{Benz et al.}
\abstract{Narrowband spikes of the decimeter type have been identified in dynamic spectrograms of Phoenix-2 of ETH Zurich and located in position with the Nan\c{c}ay Radioheliograph at the same frequency. The spike positions have been compared with the location of hard X-ray emission and the thermal flare plasma in soft X-rays and EUV lines. The decimetric spikes are found to be single sources located some 20" to 400" away from the flare site in hard or soft X-rays. In most cases there is no bright footpoint nearby. In at least two cases the spikes are near loop tops. These observations do not confirm the widely held view that the spike emission is produced by some loss-cone instability masering near the footpoints of flare loops. On the other hand, the large distance to the flare sites and the fact that these spikes are all observed in the flare decay phase make the analyzed spike sources questionable sites for the main flare electron acceleration. They possibly indicate coronal post-flare acceleration sites.
}
\date{Received ... / Accepted ...}

\maketitle
\section{Introduction}

The acceleration of large numbers of particles in flares is an old, major enigma in solar physics. Not only the question concerning the predominant mechanism is unsolved, even the location of acceleration is unclear. 

Nevertheless, over the past decade crucial observations on small relatives of flares have accumulated. They evidence that short narrowband radio bursts in meter wavelengths are signatures of the acceleration process. Benz et al. (1996) have shown that narrowband {\sl metric spikes} in general correlate with type III bursts starting at slightly lower frequencies. The metric spikes have been found by Paesold et al. (2001) to be located generally on the extension of type III trajectories and suggest a model of energy release taking place in or close to the spike sources. Modeling coronal densities (Paesold et al. 2001) and spatially resolved observations of metric spike events (Krucker et al. 1995; 1997) put the sources at altitudes of $2\times 10^{10}$cm and more. Similarly, {\sl type I} radio bursts in noise storms have been found e.g. by Raulin et al. (1994) to be accompanied by a delayed continuum radio emission and a soft X-ray brightening. The similarity of spikes and type I bursts has been emphasized previously (Benz 1985). Both noise storms and metric spike/type III bursts are not associated with regular, hard X-ray and centimeter-wave emitting flares. The short duration and the narrow bandwidth suggest a small source size and therefore a high radio brightness temperature (order of $10^{15}$ K). Only a coherent mechanism can account for the emission, but none of the published mechanisms is generally accepted. A model proposed by Benz \& Wentzel (1981) suggests type I radiation to originate in the acceleration region, where waves driven by unstable currents couple with Langmuir waves produced by accelerated electron beams. Similar processes have been proposed for narrowband spikes in the decimeter range (cf. below).

The situation is much less clear for narrowband spikes at {\sl decimeter} wavelength. The term 'narrowband, millisecond spikes' has been introduced for them referring to narrowband (few percent of the center frequency) and short (few tens of ms) peaks above 1 GHz (Droege 1977; Slottje 1978). Originally, they were reported to occur in the rise phase of centimeter radio bursts and thus to be associated with major flares. The association rate with hard X-ray flares is high (Benz \& Kane 1986; G\"udel et al. 1991). However, a delay of spike groups and hard X-ray peaks of the order of a few seconds has been noted (Aschwanden \& G\"udel 1992). Moreover, spikes have been discovered also in decimeter type IV bursts occurring after the HXR emitting phase of flares (Isliker \& Benz 1994). Contrary to their relatives at meter waves, decimetric spikes do not correlate with type III bursts. 

The emission process of decimetric spikes is highly controversial. Originally, a loss-cone instability of trapped electrons has been proposed to produce electron cyclotron maser emission at the footpoints of flare loops (Holman et al. 1980; Melrose \& Dulk 1982). To avoid the assumption of high magnetic field strength in the source, the model has been changed to emission of upper-hybrid and bernstein modes (Willes \& Robinson 1996). The scheme can interpret occasional harmonic emission in decimeter spikes (Benz \& G\"udel 1987; Krucker \& Benz 1994). Alternatively, Tajima et al. (1990) and G\"udel \& Wentzel (1993) proposed the spike sources to be in the acceleration regions of flares and to result from waves produced by the acceleration process.

Here we set out to test the predictions of the two emission models concerning the source position of narrowband decimeter spikes. In the loss-cone scenario, the spike emission is a secondary phenomenon located near the footpoints of flaring loops. Hard X-ray sources, produced by precipitating electrons have been found at footpoints of such loops (Duijveman et al. 1982). Thus the loss-cone scenario predicts {\sl (i)} spikes to occur at similar locations when seen from Earth. {\sl (ii)} As hard X-rays often have double sources, the spikes may behave similarly (as proposed by Conway \& Willes 2000). In the acceleration-region scenario on the other hand, the spikes are expected to originate from single sources at some distance from the footpoints, at higher altitude, possibly at the top of loops, and to be single sources.

Previously, Gary et al. (1991) have reported imaging observations of possible decimetric spikes at 2.8 GHz originating some 25" from the gyrosynchrotron source. An observation at 5.7 GHz of a limb flare reported by Altyntsev et al. (1995) suggests the spikes to be located within the region of gyrosynchrotron emission about 35 Mm above the photosphere. These observations did not have spectral resolution to confirm the narrowbandedness of the emission. As there are several other types of short duration coherent emission in the decimeter range, the identification of these observations as narrowband spikes remains controversial. 

Here we start out from spike events that are well identified by spectrometer observations and locate their position with simultaneous measurements by an interferometer observing at the frequencies of spike emission. The position of the radio emission is then compared with imaging observations in hard X-rays, if available. Finally, the sources are put into the context of the thermal coronal plasma and field geometry as imaged by coronal EUV lines and soft X-rays.

\section{Instruments and Observations}
\subsection{The Phoenix-2 spectrometer}
The Phoenix-2 spectrometer operated by the ETH Zurich in Bleien (Switzerland) since 1998 is continuously recording radio data from sunrise to sunset. The full Sun flux density  and circular polarization are measured by the frequency-agile 
receiver in the range from 0.1-4.0 GHz at a time resolution of 500 $\mu$s for single channel measurements. A description of the instrument can be found in Messmer et al. (1999). The data used herein were recorded at a bandwidth of 1 MHz and a time resolution of 100 ms in the frequency range from 220 to 553 MHz, chosen for collaboration with the Nan\c{c}ay Radioheliograph. In addition, the frequencies of 1415 and 2800 MHz were recorded. As Phoenix-2 polarization measurements were intermittent and partially unreliable in the observing period, the NRH data were used.
\subsection{The Nan\c{c}ay Radioheliograph (NRH)}
The 2D-imaging radioheliograph in Nan\c{c}ay (France) is a 43 element interferometer dedicated to solar observations (Kerdraon \& Delouis 1996). Up to 10 frequencies in the range from 150-450 MHz can be observed simultaneously at a maximum number of 200 images per second. The antennas are placed in two perpendicular arrays and digitally correlated in 576 channels. The Stokes parameters I and V are measured. The observing bandwidth is 700 kHz. Data for this analysis were taken at the center frequencies of 164.0, 236.6, 327.0, 410.5, and 432.0 MHz, and with a mean time resolution of 120 ms. The relative accuracy of source positions is limited by noise and digitization (pixel size). 
\subsection{The Hard X-ray Telescope}

Data from the Hard X-ray Telescope (HXT, Kosugi et al. 1991) on the Yohkoh spacecraft are used when available.
\subsection{Coronal EUV and Soft X-ray Observations }

Observations of the thermal background plasma of the corona are used to locate the spike sources and hard X-ray emission in the active region and the thermal flare plasma. Simultaneous observations in either coronal EUV lines or soft X-rays were preferred. If not available, proxies close in time were used. Yohkoh, EIT, and TRACE took the data that were finally employed. The Soft X-ray Telescope (SXT) on-board Yohkoh (Tsuneta et al. 1991) provided full and partial disk images in the 0.25 - 4 keV energy range with pixel sizes down to 2.45". Five different filters can be positioned in front of the detector array to estimate isothermal temperatures and emission measure. The Extreme ultraviolet Imaging Telescope (EIT) onboard the Solar and Heliospheric Observatory (SoHO) is a normal-incidence, multilayered mirror instrument (Delaboudini\`ere et al. 1995) having a spatial resolution of 2.6". We used mostly the wavelength band at 195 \AA\ that includes a line of Fe XII, sensitive to coronal plasma in the temperatures range 0.5 to 2.0 $\times 10^6$K. TRACE data (Handy et al. 1999) at 195 \AA\ were also used. 

\subsection{Selection of Events}
Calibrated data from observations of Phoenix-2 were searched for decimetric spike emission. Decimetric spikes are easy to distinguish from metric spikes in dynamic spectrograms. They appear in clusters of spikes over a total bandwidth with a maximum-to-minimum frequency ratio of more than a factor of two. The minimum frequency is usually higher than 300 MHz. Only spike events were selected that included at least one frequency observed by the Nan\c{c}ay interferometer to locate the radio emission. This requirement limited the choice of events to the lowest frequencies where decimetric spikes occur. 

Five events were found in the recent data. Three of them occurred on the same day, but all had different appearances in the spectrogram.\\ 
\\
{\em 2000/07/12}: Decimeter spikes appeared in similar clusters at the times around 10:46, 10:53, 11:01, and 11:04 UT in the frequency range from about 400 to beyond the maximum observing frequency at 553 MHz. The associated flare was of GOES class X1.9. The centimeter waves peaked at 10:31 - 33 UT, but lasted until 11:07 UT at 15.4 GHz (Solar and Geophysical Data). San Vito reported an H$\alpha$ flare location of N17E27. The 2800 MHz flux showed a small, gradual enhancement peaking at the time of the spike emission. \\
\\
{\em 2000/07/12}: A different group of spike clusters was apparently associated with an H$\alpha$ flare location at N17E18 peaking at 11:49 UT. The first three spike clusters, at 11:55, 11:58 and 12:01 UT have similar spectra. Spikes are observed in the range from 250 -- 500 MHz. Above that frequency a continuum emission is associated. The fourth event at 12:11 UT shows spike emission up to the limiting frequency of 553 MHz. Again, weak 2.8 GHz emission was associated. \\
\\
{\em 2000/07/12}: At 13:39 UT a cluster of spikes similar to the forth cluster in the previous event occurred. There was very little flux registered in the 2.8 GHz channel, but a simultaneous enhancement can be noticed. Contrary to the previous events, the polarization of the cluster is not complete, but amounts to 75$\pm$10 \% left circular.\\
\\
{\em 2000/07/20}: A cluster of spikes was recorded starting at 12:05 UT. A possibly related soft X-ray event had its maximum some 10 minutes before. No centimeter wave emission was reported, but the 2.8 GHz channel of the Phoenix-2 spectrometer recorded a weak, exactly simultaneous enhancement during the spike activity. The polarization of the spikes is intermediate (40$\pm$10 \% left circular). \\
\\
{\em 2000/09/23}: A centimeter burst was reported up to 8.8 GHz, peaking at 11:05 UT, the start time of the spikes.  Their circular polarization is low (0 -- 40 \% left). At 2.8 GHz the emission reached the maximum at the same time as the spikes. 

For none of the events a coincident major peak in hard X-rays or centimeter emission has been reported. We note that all selected events occurred after the main flare peak. Although this is not unusual, it is not consistent with the initial reports of narrowband spikes at decimeter waves that were made at higher frequencies (cf. Section 1). Nevertheless, the selected events show a gradual enhancement at 2.8 GHz in all cases. It is simultaneous with the spikes within seconds and may be interpreted as gyrosynchrotron emission. 

High polarization of spikes is not untypical for decimetric spikes near the center of the solar disk. Beyond about 260 arcseconds from disk center, intermediate and low polarizations have also been reported (G\"udel \& Zlobec 1991)\\

Examples of radio spectrograms of decimetric spikes are presented in Figs. 1, 2, 6, and 8. When seen at high time resolution, the individual spikes do not appear to be different. However the appearance of the cluster and the context of other radio emissions vary considerably among these examples. In Figure 1 the spikes extend over a relatively narrow range of frequencies, and a decimetric continuum extending beyond 1415 MHz was simultaneously emitted at higher frequency. Figure 2 presents an example of a very rich cluster of spikes that occurred only 16 minutes later than the event of Fig.1. Even richer was the event at 13:39 UT. Individual spikes were less an less visible in the three events on the same day.

\section{Results}

\subsection{Positions of Narrowband Spikes}
\begin{figure}
\begin{center}
\leavevmode
  \resizebox{9cm}{!}{\hspace{-0.2cm}\includegraphics{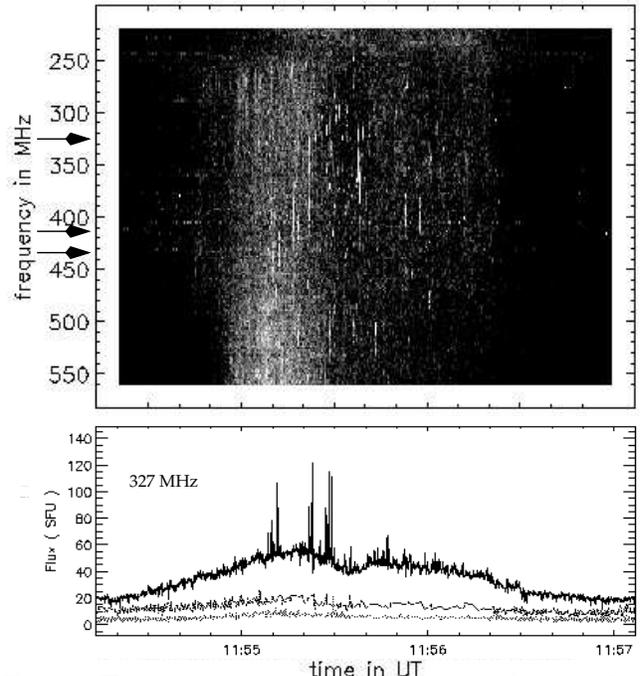}}
\end{center}
  \vskip-0.5cm
\caption[]{{\bf Top:} Spectrogram observed by Phoenix-2 on
    2000/07/12. Bright pixels correspond to enhanced flux; the 
    frequency axis is from top to bottom. The arrows
    indicate the frequencies observed by NRH. 
    {\bf Bottom:} NRH light curves 
    at 327 MHz of the three major sources present on the solar disk.}  
  \label{fig:1}
\end{figure}

Figure 1 shows the dynamic spectrum (total power, Stokes I) from 220 MHz to 553 MHz of the first spike cluster in the second event on 2000/07/12. Spikes can be found in the frequency range from 250 to 450 MHz. At higher frequencies, a broadband continuum appears simultaneously. Three radio sources were seen at 327 MHz on the solar disk at the time of the spikes (cf. time profiles in Figure 1). However, only the strongest source correlates with the spikes seen in the spectrogram. It is indicated in Fig.3.
\begin{figure}
\begin{center}
\leavevmode
  \resizebox{9.5cm}{!}{\hspace{-1.0cm}\vspace{-1.5cm}\includegraphics{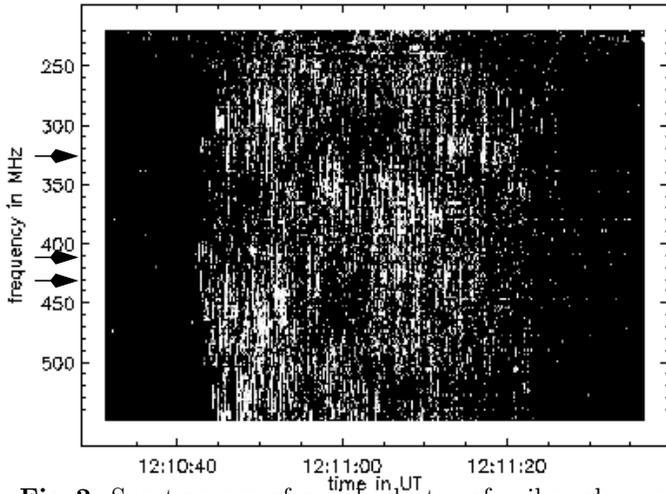}}
\end{center}
\vskip-1.0cm
  \caption[]{Spectrogram of a rich cluster of spikes observed by Phoenix-2 on 2000/07/12. The arrows indicate the frequencies observed by NRH.}
  \label{fig:2}
\end{figure}

A very different cluster of spikes is shown in Fig. 2. It occurred 15 minutes after the spikes shown in Fig. 1 and seems to be also part of the second event on 2000/07/12. The cluster contains much more spikes, which extend from about 250 MHz to beyond 550 MHz. Individual spikes have bandwidth and duration similar to the ones in Fig. 1.

\begin{figure}
\begin{center}
\leavevmode
  \resizebox{11cm}{!}{\includegraphics{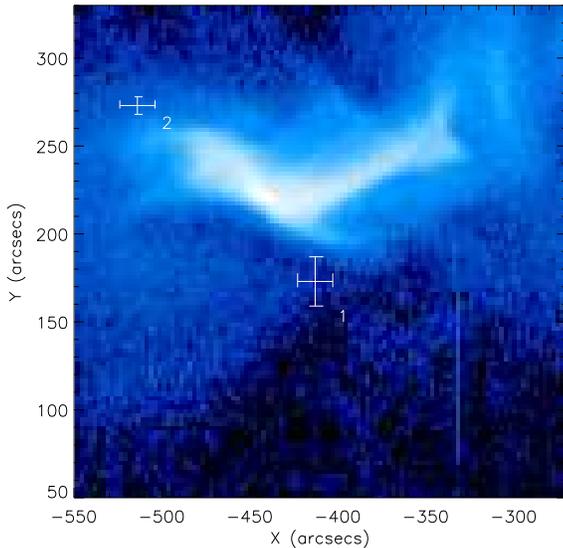}}
\end{center}
\vskip-0.5cm
  \caption[]{SXT/Yohkoh image (AlMg filter, 2.45" resolution) of the flaring
    region at 12:07:45 UT on 2000/07/12. The centroid positions of the spike sources at 327 MHz as observed by the NRH are drawn with error bars representing their scatter in time. Group 1: 11:54:40 - 11:56:15 UT (Fig.1); group 2: 12:10:40 - 12:11:20 UT (Fig.2).}
  \label{fig:3}
\end{figure}

The positions of the two spike clusters are shown in Fig. 3. The centroids of the sources at 327, 410, and 432 (only observable in group 2) are coincident within the scatter in time. The temporal scatter of the centroid indicated by the error bars varies with frequency and event. It is usually about 11", given by the pixel size used for image reconstruction, and often exceeding the statistical scatter of about 3" (depending on flux). The two clusters originate from different sources that are 140" apart. Both are not located in regions of high soft X-ray emissivity or near bright footpoints. On the other hand they are not associated with the brightest flare loops either. They rather seem to be located close to diffuse soft X-ray features, similar to metric spikes (Paesold et al. 2001, Fig. 6).

Figures 1 to 3 indicate diversity of narrowband spike events in the dynamics of the same flaring region. The positional information suggests that the radio emission originates in different loops.

\begin{figure}
\begin{center}
\leavevmode
  \resizebox{11.5cm}{!}{\includegraphics{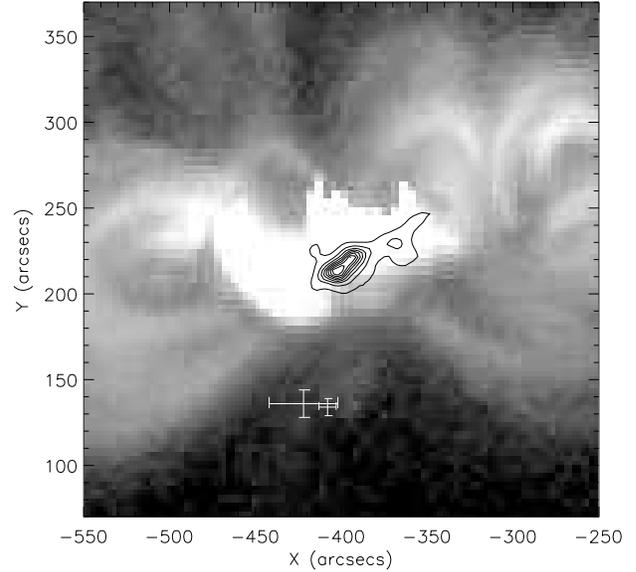}}
\end{center}
\vskip-0.3cm
  \caption[]{Spike positions in the course of the X1.9 event on 2000/07/12 superposed on an SXT/Yohkoh image (AlMg filter, 5" resolution) taken at 10:21:18 UT. The centroid positions of the spike sources at 432 MHz as observed by the NRH are drawn with error bars representing their scatter in time. Group 2 (large error bars): 10:53:20 - 10:53:40 UT; group 3 (small error bars): 11:01:00 - 11:01:10 UT. The hard X-ray intensity at 10:30:40 UT as observed by HXT/Yohkoh (M1 channel) is displayed by isophotes.}
  \label{fig:4}
\end{figure}

\subsection{Comparison with Hard X-rays}

Figure 4 shows the coincident positions of two spike groups in relation to the hard X-rays at peak flux and the thermal plasma. The first group of spikes on July 12, 2000, appeared 8 minutes after an X1.9 flare and during a time of enhanced hard X-ray emission. The spike positions are relatively stable in time and very similar in frequency. The spikes were located at 80" from the hard X-ray maximum. In fact, they were farther away from the main flare site than in the following, much smaller flare (Fig.3). The spikes originated from a region of low emissivity. The image of Fig.4 shows the spikes in apparent association with an amorphous patch of weak soft X-ray emission, possibly a faint small loop system south of the hard X-ray source. If the spikes were located at footpoints of loops, they would not be in the same loops as the ones containing the hard X-ray emitting electrons.

\begin{figure}
\begin{center}
\leavevmode
  \resizebox{12cm}{!}{\includegraphics{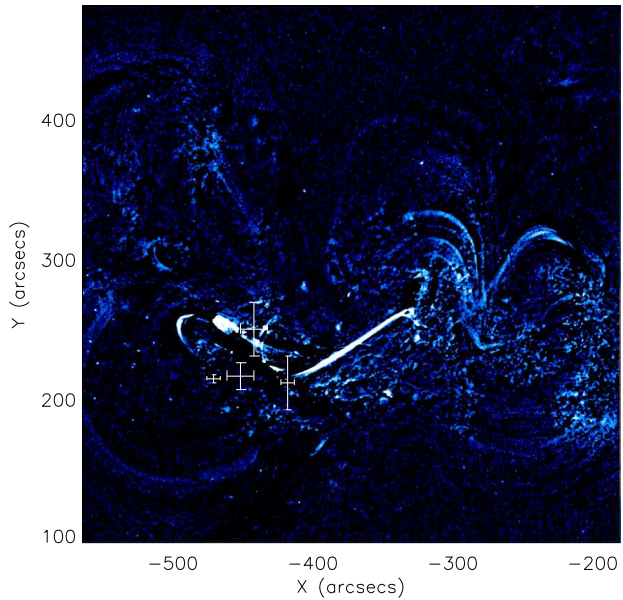}}
\end{center}
\vskip-0.3cm
  \caption[]{Centroid positions of spikes observed on 2000/07/12 13:38 - 13:40 UT at 432, 410, 327, and 236 MHz by the NRH (from left to right). They are superimposed on a TRACE image at 195 \AA\ with 1" resolution showing the difference between the image at the end of the spike event and the image just before.}
  \label{fig:5}
\end{figure}

\subsection{Comparison with Changes of the Thermal Coronal Plasma}

More structure of the active region can be seen in Fig.5 taken by TRACE during the occurrence of the third spike event on the same day. The EUV observations in Fe XII outline the thermal plasma in the temperature range 0.5 - 2.0 MK. Images at intervals of 40 seconds are available. Figure 5 displays a difference image where regions that increased their emission during the spike event are shown bright. A brightening indicates an increase in the emission measure of plasma in the sensitive range by heating and filling flux tubes with additional material or cooling of material previously hotter than 2 MK. In both cases it points to a location of rapid change.

Within the two minutes duration of the spike event, two major brightenings occurred: a thin, long loop extending to the Northwest and a loop extending to the Northeast, possibly shortened by projection. The spike centroids are in-between, overlaying approximately a bright soft X-ray structure recorded by SXT a few minutes before the event.

The event of Fig.5 is the only one in which the centroid positions at the various frequencies do not coincide. The sequence of positions is according to frequency, indicating a real effect. The spikes form a rich cluster in the spectrogram extending beyond the spectrum observed by Phoenix-2. In the low frequency part, drifting chains of spikes can be seen. As previously noted, this cluster differs from the previous ones in its lower circular polarization. It cannot be excluded that this cluster represents a different type of emission.

\subsection{Narrowband Spikes and Other Radio Observations}

\begin{figure}
\begin{center}
\leavevmode
  \resizebox{9cm}{!}{\hspace{-0.8cm}\vspace{-0.7cm}\includegraphics{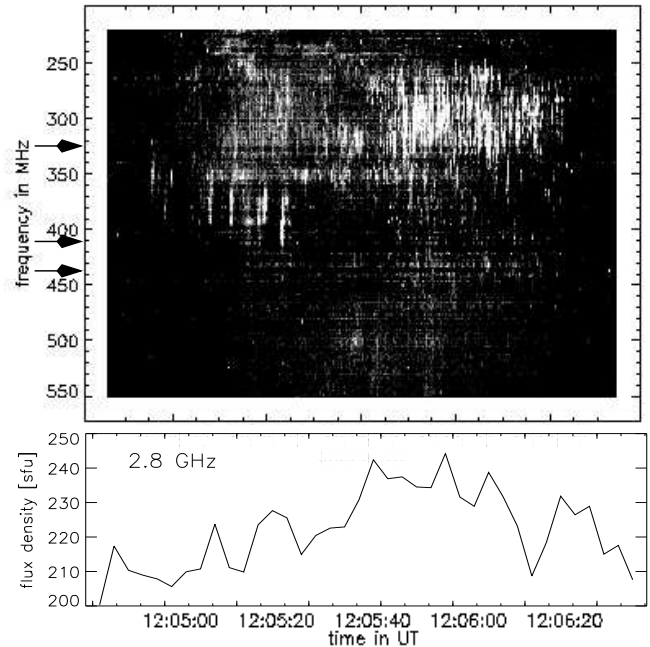}}
\end{center}
\vskip-0.3cm
  \caption[]{Spectrogram ({\sl top:}) and 2.8 GHz flux ({\sl bottom}) observed by Phoenix-2 on 2000/07/20. Note spike cluster extending from 260 MHz to 420 MHz. Regular type III bursts occur early in the event in the frequency range 330 - 430 MHz.}
  \label{fig:6}
\end{figure}

Figure 6 presents a rich cluster of spikes at exceptionally low frequencies. However, the latter did not exclude an associated 2.8 GHz emission that correlates well with the spikes and peaks at the maximum of the spike activity.

\begin{figure}
\begin{center}
\leavevmode
  \resizebox{8cm}{!}{\includegraphics{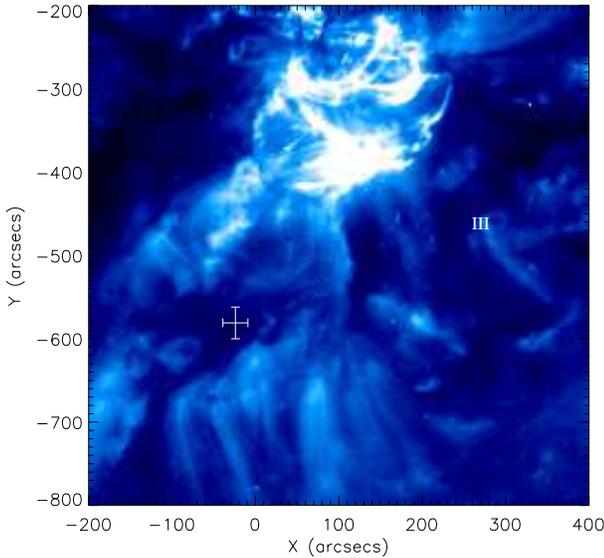}}
\end{center}
\vskip-0.3cm
  \caption[]{Centroid positions of spikes (error bars) on 2000/07/20 (cf. Fig.6) at 327 and 410 MHz (coinciding). The position of type III bursts (III) at 410 MHz is also indicated. They are superimposed on an EIT image observed at 12:05:11 UT in 195 \AA. Its spatial resolution is 2.62".}
  \label{fig:7}
\end{figure}

The spike centroids observed at two frequencies coincide within the error bars defined by the positional scatter of the sources in time. Figure 7 shows the position in a region at a large distance from the main active region in an area of very low Fe XII emissivity. All nearby loops appear to bend away from this position. The average position is more than 30" apart from the nearest structure discernible at 195 \AA .

In addition, a group of more than 15 type III bursts occurred between 12:04:55 and 12:05:25 UT early in the event, but simultaneous with weak spike activity at lower frequency. The bursts have negative frequency drifts, indicative of electron beams propagating upward in the corona. The type III position at 410 MHz is about 320" Northwest of the spikes. Though they occur in the same active region at nearly the same time, the relation between the two types of radio emission is unclear. 

\begin{figure}
\begin{center}
\leavevmode
  \resizebox{9cm}{!}{\hspace{-0.7cm}\includegraphics{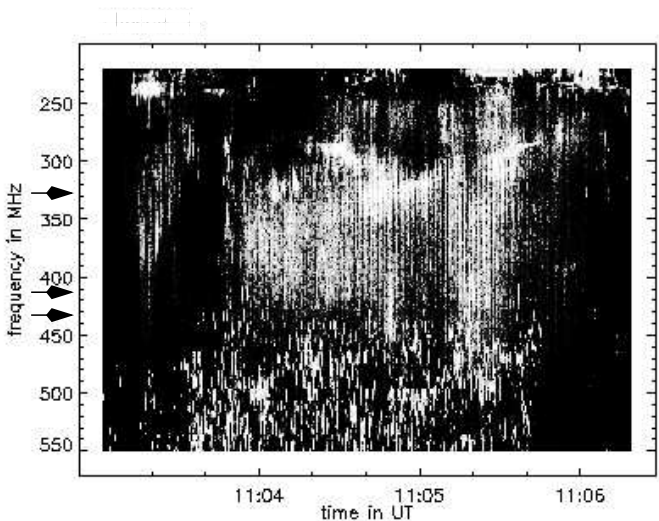}}
\end{center}
\vskip-0.3cm
  \caption[]{Spectrogram of radio emission observed by Phoenix-2 on 2000/09/23. A variety of types of emission is discernible: decimetric spikes extending from 400 MHz to beyond 550 MHz; drifting pulsations 270 - 430 MHz in the early phase until 11:03:50 UT, after that a very regular train of more than hundred pulsations with a period of 0.98 s, and meter wave activity throughout the event.}
  \label{fig:8}
\end{figure}

The spikes on 2000/09/23 (Fig.8) were at relatively high frequencies. Nevertheless, a few spikes did occur at the highest frequencies observable by the NRH. The source positions could reliably be determined from the comparison of identified spikes in the spectrogram and the time profile of the various sources observed on single frequencies. 

The positions of the spikes at 410 and 432 MHz are identical within the scatter in time. They are located (in projection) on the top of a bright soft X-ray loop (Fig.9). The associated loop has a height of about 300" (a third of a solar radius in projection). Loops can be seen further out in overexposed images but are much fainter.

More than a hundred pulsations with regular periods of 0.95 s can be seen in the spectrum (Fig.8) from 310 to 440 MHz. Spikes and pulsations overlap between 400 and 440 MHz They can easily be separated in the image. The sources of the pulsations at 327 and 410 MHz are found Northeast of the spikes, separated by 60" - 235". Furthermore, the positions of the pulsations at different frequencies are significantly displaced from each other, even when measured at the same time. In none of them are nearby loop tops visible. The positions rather correspond (in projection) to the middle part of very large, bright loops possibly of the same loop system.

Decimetric pulsations are generally interpreted by loss-cone emission of trapped energetic electrons (review by Kuijpers 1980). The hypothesis suggests that most emission occurs at intermediate loop heights where the upper hybrid frequency matches a harmonic of the electron gyrofrequency. This scenario is supported by the present observation.

\begin{figure}
\begin{center}
\leavevmode
  \resizebox{8cm}{!}{\includegraphics{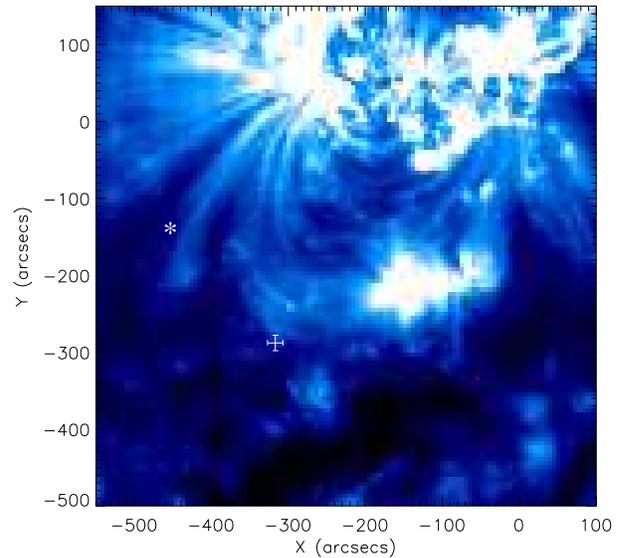}}
\end{center}
\vskip-0.3cm
 \caption[]{Centroid positions of the spike cluster at 420 and 432 MHz (coincident, same error bars) on 2000/09/23 observed with NRH. Also shown is the position of pulsations at 410 MHz (asterisk). They are superimposed on an EIT image observed at 11:12:10 UT in 195 \AA\ with a spatial resolution of 2.65".}
  \label{fig:9}
\end{figure}

\section{Discussion and Conclusions}

The requirement to choose decimetric spike events at low frequency to be observable with the NRH has produced a selection of five events that all occurred in the flare decay phase. The analysis indicates that 

1. Generally, spike sources are not found at footpoints of flare loops as indicated by soft X-ray or coronal EUV emissions. Only one case (Fig.7) seems to be nearer to footpoints than loop tops, but the footpoints are not those of the main flare loops. There are two clear cases (Figs. 4 and 9) and three possible cases (two in Fig. 3 and one in Fig. 5) of a coincidence with a loop top in projection. 

2. Consistent with this, we find the spike sources at very distant locations from the site of hard X-ray emission (Fig.4). The angular distance from the main soft X-ray or EUV loops is 20" to 400". Only in one case (Fig.5), does the spike sources coincide with strong soft X-ray emission, which may be a projection effect.

3. None of the spike sources was found to be double at a given time.

4. The spike sources, if measured at different frequencies, appear to be located at the same site within the intrinsic scatter they have in time (Figs. 4, 7, and 9). There is however a clear exception (Fig.5), where the positions are spread according to frequency. 

5. There may be more than one source of spike emission in the course of an event. In Fig. 3 a case is shown where two clusters separated by 16 minutes in time originate from widely different places in the same active region.

6. Other radio emissions at the same frequency may originate at far away positions, such as observed for simultaneous type III bursts (Fig. 6) and decimetric pulsations (Fig. 8).

The fact that the spike sources are found far away from the major flare site may be related to the low frequency events we have selected. Combined with their occurrence after the main flare phase, they are unlikely to be the main sites of flare electron acceleration. Nevertheless, these spike sources cannot be excluded as secondary acceleration sites as they have all been associated with 2.8 GHz radiation of potential gyrosynchrotron origin. Thus we suggest that the observed spikes are related to the rearrangement of the coronal magnetic field after the main flare at lower altitude. They may be indicative of post-flare high-corona acceleration sites.

Narrowband spikes at decimetric frequencies have been reported in the flare rise phase and main phase, even at low frequencies (Benz 1985). It will be interesting to investigate the above findings on the location of spikes earlier in the flare and at higher decimetric frequencies. More sensitive and precise information on hard X-rays are urgently needed. Such a work is well suited for the pending HESSI mission.

\begin{acknowledgements}
We thank Christian Monstein, Michael Arnold and Peter Messmer (ETH Zurich) 
for helping to run the Phoenix-2 observations. The evaluation of the 
Phoenix spectrograms was made using software developed by A. Csillaghy. 
The Swiss National Science Foundation financially supports the work at ETH Zurich (grant No. 2000-061559.00). The  Nan\c{c}ay Radio Observatory is funded by the French Ministry of Education, the CNRS and the R\'egion Centre. We acknowledge data from Yohkoh/HXT, Yohkoh/SXT, EIT and TRACE. The Yohkoh mission is operated by ISAS and an international cooperation including NASA and SERC. CNES, NASA, and the Belgian SPPS funded the EIT instrument on ESA's SoHO spacecraft. TRACE is a NASA small explorer mission.
\end{acknowledgements}


\begin{thebibliography}{} 
\bibitem[]{} Altyntsev, A.T., Grechnev, G.N., Zubkova, G.N., Kardapolova, N.N., Lesovoi, S.V., Rosenraukh, Yu.M., Treskov, T.A.: 1995, A\&A 303, 249
\bibitem[]{} Aschwanden, M.J., G\"udel, M., 1992, ApJ 401, 736
\bibitem[]{} Benz, A.O., 1985, Solar Phys. 104, 99
\bibitem[]{} Benz, A.O., Csillaghy, A., Aschwanden, M.J., 1996, A\&A 309, 2 291
\bibitem[]{} Benz, A.O., Zlobec, P., Jaeggi, M., 1982, A\&A 109, 305
\bibitem[]{} Benz, A.O., Kane, S.R., 1986, Solar Phys. 104, 179
\bibitem[]{} Benz, A.O., G\"udel, M., 1987, Solar Phys. 111, 175
\bibitem[]{} Benz, A. O.; Wentzel, D. G., 1981, Solar Phys. 94, 100
\bibitem[]{} Conway, A. J.; Willes, A. J., 2000, A\&A 355, 751
\bibitem[]{} Delaboudini\`ere, J.-P. et al., 1995, Solar Phys. 162, 291 
\bibitem[]{} Droege, F., 1977, A\&A 57, 285
\bibitem[]{} Duijveman, A.; Hoyng, P.; Machado, M. E., 1982, Solar Phys. 81, 137
\bibitem[]{} Gary, D.E., Hurford, D.J., Flees, D.J.: 1991 ApJ 369, 255
\bibitem[]{} G\"udel, M., Aschwanden, M.J., Benz, A.O., 1991, A\&A 251,285
\bibitem[]{} G\"udel, M., Zlobec, P., 1991, A\&A 245, 299
\bibitem[]{} G\"udel, M., Wentzel, D.G., 1993, ApJ 415, 750
\bibitem[]{} Handy, B., et al. 1999, Solar Phys., 187, 229
\bibitem[]{} Holman, G.D., Eichler, D., Kundu, M.R., 1980, In: Radio Physics of the Sun, Kundu M.R. and Gergeley T.E. (eds.), IAU Symp., Vol. 86, 457 
\bibitem[]{} Isliker, H.; Benz, A. O., 1994, A\&AS  104, 145
\bibitem[]{} Kerdraon, A., Delouis, J.-M., 1996, The Nan\c{c}ay Radioheliograph. In: Coronal Physics from Radio and Space Observations, Trottet G. (ed.), Springer Verlag, Berlin, 192
\bibitem[]{} Kosugi T., Makishima K., Murakami T., et al., 1991, Solar Phys. 136, 17
\bibitem[]{} Krucker, S., Aschwanden, M.J., Bastian, T.S., Benz, A.O., 1995, A\&A 302, 551 
\bibitem[]{} Krucker, S., Benz, A.O., 1994, A\&A 285, 1038
\bibitem[]{} Krucker, S., Benz, A.O., Aschwanden, M.J., 1997, A\&A 317, 569
\bibitem[]{} Kuijpers, J., 1980, in: Radio Physics of the Sun; Proceedings IAU Symposium Nr.86 (eds.M.R.Kundu and T.E.Gergely), D. Reidel Publishing, 341
\bibitem[]{} Melrose, D.B., Dulk, G.A., 1982, ApJ 447, 844
\bibitem[]{} Messmer, P., Benz, A.O., Monstein, C., 1999, Solar Phys. 187 (2), 335
\bibitem[]{} Paesold, G., Benz, A.O., Klein, K.-L., Vilmer N., 2001, A\&A 371, 333 
\bibitem[]{} Raulin, J. P., Klein, K.-L., 1994, A\&A  281, 536 
\bibitem[]{} Slottje, C., 1978, Nature 275, 520
\bibitem[]{} Tajima, T., Benz, A.O., Thaker, M., Leboeuf, J.N., 1990, ApJ 353, 666
\bibitem[]{} Tsuneta, S. et al. 1991, ApJ 459, 342
\bibitem[]{} Willes, A. J.; Robinson, P. A., 1996, ApJ 467, 465  
\bibitem[]{} Wright, C.S., 1980, Astronomical Society of Australia,
Proceedings 4(1), 62
\end{thebibliography}
\end{document}